\documentclass[aps,prl,amsmath,lengthcheck,superscriptaddress]{revtex4-2}

\usepackage{graphicx}\graphicspath{ {figures/} }
\usepackage{hyperref}
\hypersetup{colorlinks,allcolors=blue,breaklinks}

\newcommand{\be}{\begin{equation}}
\newcommand{\ee}{\end{equation}}
\newcommand{\ba}{\begin{align}}
\newcommand{\ea}{\end{align}}
\newcommand{\bi}{\begin{itemize}}
\newcommand{\ei}{\end{itemize}}

\newcommand{\la}{\left\langle}
\newcommand{\ra}{\right\rangle}
\newcommand{\pd}{\partial}

\newcommand{\bla}{bla\\bla\\bla\\bla\\bla}

\newcommand{\mb}[1]{\mbox{\boldmath$#1$}}
\newcommand{\mc}[1]{\mathcal{#1}}

\begin{document}

\title{Analytical solution for optimal protocols of weak drivings}

\author{Pierre Naz\'e}
\email{pierre.naze@icen.ufpa.br}

\affiliation{\it Universidade Federal do Par\'a, Faculdade de F\'isica, ICEN,
Av. Augusto Correa, 1, Guam\'a, 66075-110, Bel\'em, Par\'a, Brazil}
\date{\today}

\begin{abstract}

One of the main objectives of science is the recognition of a general pattern in a particular phenomenon in some particular regime. In this work, this is achieved with the analytical expression for the optimal protocol that minimizes the thermodynamic work and its variance for finite-time, isothermal, and weak processes. The method that solves the Euler-Lagrange integral equation is quite general and depends only on the time-reversal symmetry of the optimal protocol, which is proven generically for the regime considered. The solution is composed of a straight line with jumps at the boundaries and impulse-like terms. Already known results are deduced, and many new examples are solved corroborating this pattern. Slowly-varying and sudden cases are deduced in their appropriate asymptotic limits. Comparison with numerical procedures is limited by the nonavailability of the present methods of the literature to produce solutions in the space of distributions.

\end{abstract}

\maketitle

\section{Introduction}
\label{sec:introduction}

The usual path of scientific investigation is the accumulation of facts from nature and a constant attempt to recognize patterns in them. This is done until the point where a breakthrough is proposed with a general theoretical formulation that encompasses all the previous results. Sometimes, the topic of research is so difficult and complex that achieving this last step is out of consideration.

Optimization problems, with its plethora of particular study cases, are in this kind of classification \cite{gelfand2000,kirk2004,boyd2004,weise2009,ruszczynski2011,aurell2011,sivak2012,zulkowski2015,sivak2016,large2018}. In particular, the problem of finding the optimal protocols to minimize the thermodynamic work and its variance for isothermal, finite-time, and weak drivings shares the same quality~\cite{naze2022optimal,naze2023optimal}. Indeed, the solution of the Euler-Lagrange integral equation associated, given by a Fredholm integral equation of first kind and constant limits of integration, highly depends on the kernel proposed~\cite{polyanin2008handbook}, becoming a case-to-case study. Therefore, studying particular examples or building numerical procedures seems to be a better course than finding a general aspect that unifies analytically this phenomenon in this particular regime~\cite{acconcia2015b,deffner2018,kamizaki2022}. Notorious exceptions however have been found in the context of rapidly driven systems~\cite{blaber2021steps,rolandi2023optimal}, and geometric thermodynamics applied to Fokker-Planck formalism~\cite{ito2024geometric}.

The present work goes in the opposite direction and aims to solve the Euler-Lagrange integral equation of the referred problem in a general way. The method proposed is quite general and depends only on the time-reversal symmetric property of the optimal protocol~\cite{naze2022optimal,loos2024universal}, which is proven generically. I find that the optimal protocol, for any reasonable physical system in the regime proposed, is composed of a straight line with jumps at the beginning and end of the process, and a series of Dirac deltas and their derivatives. Also, it depends exclusively on the ratio between the switching time of the process and the relaxation timescale of the system. To illustrate the method, I present many examples, from Brownian motion to the Ising chain, recovering already known results~\cite{schmiedl2007optimal,gomez2008optimal} and finding the pattern in new ones. The slowly-varying and sudden cases are calculated in their appropriate asymptotic limits. At the end, a brief discussion is made about the limits of the available methods present in the literature to produce numerical solutions in the space of distributions.

\section{Preliminaries}
\label{sec:preliminaries}

I start by defining notations and developing the main concepts to be used in this work. This section is based on the technical introductory section of Ref.~\cite{naze2023optimal}.

Consider a classical system with a Hamiltonian $\mc{H}(\mb{z}(\mb{z_0},t)),\lambda(t))$, where $\mb{z}(\mb{z_0},t)$ is a point in the phase space $\Gamma$ evolved from the initial point $\mb{z_0}$ until time $t$, with $\lambda(t)$ being a time-dependent external parameter. During a switching time $\tau$, the external parameter is changed from $\lambda_0$ to $\lambda_0+\delta\lambda$, with the system being in contact with a heat bath of temperature $\beta\equiv {(k_B T)}^{-1}$, where $k_B$ is Boltzmann's constant. The average work performed on the system during this interval of time is
\be
\overline{W}(\tau) \equiv \int_0^\tau \la\overline{\pd_{\lambda}\mc{H}_T}(t)\ra_0\dot{\lambda}(t)dt,
\label{eq:work}
\ee
where $\mc{H}_T$ is the total Hamiltonian composed of $\mathcal{H}$ and that of the heat bath, $\partial_\lambda$ is the partial derivative in respect to $\lambda$ and the superscripted dot the total time derivative. The generalized force $\la\overline{\pd_{\lambda}\mc{H}_T}\ra_0$ is calculated using the averaging $\overline{\cdot}$ over the stochastic path and the averaging $\langle\cdot\rangle_0$ over the initial canonical ensemble. The external parameter can be expressed as
\be
\lambda(t) = \lambda_0+g(t)\delta\lambda,
\label{eq:ExternalParameter}
\ee
where to satisfy the initial conditions of the external parameter the protocol $g(t)$ must satisfy the following boundary conditions
$g(0)=0$ and $g(\tau)=1$.

Linear-response theory aims to express average quantities until the first-order of some perturbation parameter considering how this perturbation affects the observable to be averaged and the probabilistic distribution \cite{kubo2012}. In our case, we consider that the parameter does not considerably change during the process, $|g(t)\delta\lambda/\lambda_0|\ll 1$, for all $t\in[0,\tau]$ and $\lambda_0\neq 0$. The generalized force can be approximated until the first order as
\begin{equation}
\begin{split}
\la\overline{\pd_{\lambda}\mc{H}_T}(t)\ra_0 =&\, \la\pd_{\lambda}\mc{H}_T\ra_0-\delta\lambda\widetilde{\Psi}_0 g(t)\\
&+\delta\lambda\int_0^t \Psi_0(t-t')\dot{g}(t')dt',
\label{eq:genforce-relax}
\end{split}
\end{equation}
where 
\be
\Psi_0(t) = \beta\la\pd_\lambda\mc{H}_T(0)\overline{\pd_\lambda\mc{H}_T}(t)\ra_0-\mc{C}
\ee 
is the relaxation function and $\widetilde{\Psi}_0\equiv \Psi_0(0)-\la\pd_{\lambda\lambda}^2\mc{H}_T\ra_0$ \cite{kubo2012}. The constant $\mc{C}$ is calculated to vanish the relaxation function for long times \cite{kubo2012}. In general, the relaxation function can always be proposed in a phenomenological way, since it is an even function in time and presents a positive Fourier transform~\cite{naze2020}. Also, the relaxation timescale of the system is defined as
\be
\tau_R=\int_0^\infty \frac{\Psi_0(t)}{\Psi_0(0)}dt.
\ee
Combining Eqs.~\eqref{eq:work} and \eqref{eq:genforce-relax}, the average work performed at the linear response of the generalized force is
\begin{equation}
\begin{split}
\overline{W}(\tau) = &\, \delta\lambda\la\pd_{\lambda}\mc{H}_T\ra_0-\frac{\delta\lambda^2}{2}\widetilde{\Psi}_0\\
&+\delta\lambda^2 \int_0^\tau\int_0^t \Psi_0(t-t')\dot{g}(t')\dot{g}(t)dt'dt.
\label{eq:work2}
\end{split}
\end{equation}
Observe that the double integral on Eq.~\eqref{eq:work2} vanishes for long switching times \cite{naze2020}, which indicates that the other terms are the contribution of the difference of Helmholtz's free energy. The irreversible work $W_{\rm irr}$ is therefore
\begin{equation}
\begin{split}
W_{\rm irr}(\tau) = \frac{\delta\lambda^2}{2} \int_0^\tau\int_0^\tau \Psi_0(t-t')\dot{g}(t')\dot{g}(t)dt'dt,
\label{eq:wirrLR}
\end{split}
\end{equation}
where the symmetric property of the relaxation function was used \cite{kubo2012}. The regime where such expression holds is the finite-time and weak processes, where the ratio $\delta\lambda/\lambda_0\ll 1$, while $\tau_R/\tau$ is arbitrary.

Consider the irreversible work rewritten in terms of the protocols $g(t)$ instead of its derivative
\be
\begin{split}
    W_{\rm irr}&(\tau) = \frac{\delta\lambda^2}{2}\Psi_0(0)+\delta\lambda^2\int_0^\tau \dot{\Psi}_0(\tau-t)g(t)dt\\&-\frac{\delta\lambda^2}{2}\int_0^\tau\int_0^\tau \ddot{\Psi}_0(t-t')g(t)g(t')dt dt'.
\end{split}
\ee
Using calculus of variations, one can derive the Euler-Lagrange equation that furnishes the optimal protocol $g^*(t)$ of the system that will minimize the irreversible work \cite{naze2022optimal}
\be
\int_0^\tau \ddot{\Psi}_0(t-t')g^*(t')dt' = \dot{\Psi}_0(\tau-t).
\label{eq:eleq}
\ee
In particular, the optimal irreversible work will be \cite{naze2022optimal}
\be
W_{\rm irr}^*(\tau) = \frac{\delta\lambda^2}{2}\Psi_0(0)+\frac{\delta\lambda^2}{2}\int_0^\tau \dot{\Psi}_0(\tau-t)g^*(t)dt.
\ee
Also, the Euler-Lagrange equation \eqref{eq:eleq} furnishes the optimal protocol that minimizes the variance of the work, which is given by \cite{naze2023optimal}
\be
\frac{\beta}{2}\sigma_W^2(\tau)=W_{\rm irr}(\tau).
\label{eq:fdr}
\ee
In this case, its optimal value is
\be
\sigma_{\rm W}^{2^*}(\tau) = \frac{\delta\lambda^2}{\beta}\Psi_0(0)+\frac{\delta\lambda^2}{\beta}\int_0^\tau \dot{\Psi}_0(\tau-t)g^*(t)dt.
\ee

The objective of this work is to solve the Euler-Lagrange integral equation \eqref{eq:eleq} for any kind of relaxation function of a system performing an isothermal process. Mathematically speaking, I will present a new method to solve Fredholm integral equations of the first type with constant limits of integration for symmetric kernels. Such a method relies only on the time-reversal symmetry property of the solution~\cite{naze2022optimal,loos2024universal}. In the next Section, I show that optimal protocols are always time-reversal symmetric.

\section{Time-reversal symmetry}

The demonstration in the next section will be based on the assumption that the optimal protocol is time-reversal symmetric. In Ref.~\cite{naze2022optimal} it was only proved that if an optimal protocol exists, its time-reversal counterpart is an optimal protocol too. This, however, does not imply that they are equal to each other. Also, Ref.~\cite{loos2024universal} guarantees the time-reversibility of the optimal protocol only for Langevin dynamics with linear forces. I am going to prove now that an optimal protocol in weakly driven processes is always time-reversal symmetric.

Consider that $g^*(t)$ is an optimal protocol, that is, it is a solution for the Euler-Lagrange integral equation
\be
\int_0^\tau \ddot{\Psi}_0(t-t')g^*(t')dt' = \dot{\Psi}_0(\tau-t).
\label{eq:eleq1}
\ee
According to Ref.~\cite{naze2022optimal}, its time-reversal counterpart $1-g^*(\tau-t)$ satisfies the same equation as well
\be
\int_0^\tau \ddot{\Psi}_0(t-t')(1-g^*(\tau-t'))dt' = \dot{\Psi}_0(\tau-t).
\label{eq:eleq2}
\ee
Subtracting Eq.~\eqref{eq:eleq2} from Eq.~\eqref{eq:eleq1}, one has
\be
\int_0^\tau \ddot{\Psi}_0(t-t')(g^*(t')-(1-g^*(\tau-t')))dt' = 0.
\label{eq:eleq3}
\ee
Multiplying then Eq.~\eqref{eq:eleq3} by $g^*(t)-(1-g^*(\tau-t))$ and integrating $t$ in the interval $[0,\tau]$, one arrives at
\begin{widetext}
\be
\int_0^\tau\int_0^\tau \ddot{\Psi}_0(t-t')[g^*(t)-(1-g^*(\tau-t))][g^*(t')-(1-g^*(\tau-t'))]dtdt' = 0.
\label{eq:eleq4}
\ee
\end{widetext}
Observe that $\ddot{\Psi}_0(t-t')$ is a symmetric negative-definite kernel since its Fourier transform is negative, which implies by Bochner's theorem that the double integral in Eq.~\eqref{eq:eleq4} is negative for any non-null protocol chosen~\cite{feller1991}. In particular, since the double integral is null for this case for all $\tau$, the combination of optimal protocols must be null as well since the dimension of the null space of such integral operator is zero. In this manner 
\be
g^*(t)=1-g^*(\tau-t),
\ee
so every optimal protocol is necessarily time-reversal symmetric. We are able now to find the analytical solution.

\section{Exact solution}

To derive the solution of Eq.~\eqref{eq:eleq}, I consider the time-reversal symmetry of the optimal protocol
\be
g^*(t)=1-g^*(\tau-t).
\label{eq:symmetricprotocol}
\ee
First, I open the left-hand side of Eq.~\eqref{eq:eleq} into two integrals
\be
\int_0^t\ddot{\Psi}_0(t-t')g^*(t')dt'+\int_t^\tau\ddot{\Psi}_0(t-t')g^*(t')dt' = \dot{\Psi}_0(\tau-t).
\label{eq:aux1}
\ee
Using the symmetric property~\eqref{eq:symmetricprotocol} in the second term of the left-hand side of Eq.~\eqref{eq:aux1}, one has
\be
\int_0^t\ddot{\Psi}_0(t-t')g^*(t')dt'=\int_0^{\tau-t}\ddot{\Psi}_0(\tau-t-t')g^*(t')dt'.
\label{eq:aux2}
\ee
Therefore, the right-hand side of Eq.~\eqref{eq:aux2} must be equal to a symmetric function $h(t)$, such that, $h(t)=h(\tau-t)$. To find $h(t)$, consider a function $g_0(t)$, such that
\be
\int_0^t\ddot{\Psi}_0(t-t')g_0(t')dt'=C_0,
\ee
where $C_0$ is a constant in time. By applying the convolution theorem, one has
\be
g_0(t) = \mathcal{L}^{-1}_s\left\{\frac{C_0}{s \mathcal{L}_t\{\ddot{\Psi}_0(t)\}(s)}\right\}(t),
\ee
where $\mathcal{L}_t\{\cdot\}$ and $\mathcal{L}_s^{-1}\{\cdot\}$ are respectively the Laplace and inverse Laplace transforms. Assuming that $\mathcal{L}_t\{\dot{\Psi}_0(t)\}(s)$ can be expressed as a Taylor series in $s=0$, one has after applying Horner-Ruffini polynomial division method~\cite{hrmethod}
\be
\frac{1}{s\mathcal{L}_t\{\ddot{\Psi}_0(t)\}(s)}=\frac{1}{s^2\mathcal{L}_t\{\dot{\Psi}_0(t)\}(s)}=-\sum_{n=-2}^\infty a_n s^n.
\ee
The function $g_0(t)$ will be
\be
g_0(t) = -C_0\left(a_{-2} t + a_{-1} + \sum_{n=0}^{\infty}a_n \delta^{(n)}(t)\right),
\label{eq:g0solution}
\ee
where $\delta^{(n)}(t)$ is the $n$-th derivative of the Dirac delta. I demand the constant $C_0$ to be equal to a number where it holds the time-reversal symmetry
\be
g_0(t)+C_0 \sum_{n=0}^{\infty}a_n \delta^{(n)}(\tau-t)=1-g_0(\tau-t)-C_0 \sum_{n=0}^{\infty}a_n \delta^{(n)}(t).
\label{eq:symmetry}
\ee
Since the Dirac deltas and their derivatives cancel out, the constant becomes
\be
C_0 = -\frac{1}{a_{-2} \tau + 2 a_{-1}}.
\label{eq:c0}
\ee
Since $C_0$ does not depend on time, $g_0(t)$ is a solution to the Euler-Lagrange integral equation. To construct a visual symmetric representation, I consider
\be
\begin{split}
g^*(t) = \frac{1}{2}\left(g_0(t)+1-g_0(\tau-t)-\right.\\
\left.C_0\sum_{n=0}^{\infty}a_n (\delta^{(n)}(t)-\delta^{(n)}(\tau-t))\right),
\label{eq:gopt}
\end{split}
\ee
which after the substitution of Eqs.~\eqref{eq:g0solution} and~\eqref{eq:c0} becomes
\be
g^*(t) = \frac{a_{-2}t+a_{-1}}{a_{-2}\tau+2 a_{-1}}+\sum_{n=0}^{\infty}\frac{a_n (\delta^{(n)}(t)-\delta^{(n)}(\tau-t))}{a_{-2}\tau+2 a_{-1}}.
\label{eq:maingopt}
\ee
Equation~\eqref{eq:maingopt} is the main result of this work. It expresses that, for any physical system, the pattern of the optimal protocol is the same. Also, as expected, the optimal protocol is very related to the relaxation function of the system, according to the definitions of the terms $a_n$ that compose the distribution part. 

Such a result was derived for only one parameter. A derivation to multi-parameter control case requires a suitable formulation of the Euler-Lagrange equation problem, which probably would imply a set of integral equations to be solved. Last but not least, the uniqueness of such a solution is not guaranteed by the presented deduction. Indeed, using different symmetric functions $h(t)$ may prompt new solutions.

\section{Examples}

To illustrate the consistency of the method, I find the optimal protocols of physical systems performing isothermal processes by using the procedure above. These examples were chosen requiring only that the Fourier transforms of their relaxation functions are positive, which makes valid the Second Law of Thermodynamics for each one of them~\cite{naze2020}. Without loss of generality, I assume $\Psi_0(0)=1$.

\subsection{Overdamped Brownian motion}

I consider in this example a white noise overdamped Brownian motion subjected to a time-dependent harmonic potential, with the mass of the system equal to one, $\gamma$ as a damping coefficient and $\omega_0$ as the natural frequency of the potential. The relaxation function for both moving laser and stiffening traps~\cite{naze2022optimal} are given by
\be
\Psi_0(t)=\exp{\left(-\frac{|t|}{\tau_R}\right)},
\ee
where $\tau_R$ is the relaxation timescale of each case. By applying the method, the terms $a_n$ will be
\be
a_{-2}=1,\quad a_{-1}=\tau_R,\quad a_n= 0,\,n\ge 0.
\ee
Therefore
\be
g^*(t)=\frac{t+\tau_R}{\tau+2\tau_R},
\ee
which was first calculated by Schmiedl and Seifert~\cite{schmiedl2007optimal} for the full dynamics of the moving laser trap, but which is identical to the linear-response regime~\cite{naze2022optimal}. The stiffening trap case was calculated in Ref.~\cite{naze2022optimal}.

\subsection{Underdamped Brownian motion}

I consider here a white noise overdamped Brownian motion subjected to a time-dependent harmonic potential, with $m$ as the mass of the particle, $\gamma$ as a damping coefficient, and $\omega_0$ as the natural frequency of the potential. 

\subsubsection{Moving laser trap}

The relaxation function for moving laser trap is given by~\cite{deffner2018}
\be
\Psi_0(t)=\exp{\left(-\frac{\gamma}{2}|t|\right)}\left(\cos{\omega t}+\frac{\gamma}{2\omega}\sin{\omega |t|}\right),
\ee
where $\omega=\sqrt{\omega_0^2-\gamma^2/4}$. By applying the method, the terms $a_n$ will be
\be
a_{-2}=1,\quad a_{-1}=\tau_R=\gamma/\omega_0^2,\quad a_0= \tau_R/\gamma,\quad a_n=0,
\ee
for $n\ge 1$. Therefore, the optimal protocol will be
\be
g^*(t)=\frac{t+\tau_R}{\tau+2\tau_R}+\frac{\tau_R(\delta(t)-\delta(\tau-t))}{\gamma(\tau+2\tau_R)},
\ee
where $\tau_R$ is the relaxation timescale of the system. This result was first calculated by Gomez-Marin and co-authors for the full dynamics~\cite{gomez2008optimal}, which is identical to the linear-response regime~\cite{deffner2018}. 

\subsubsection{Stiffening trap}

For the stiffening trap case, the relaxation function is~\cite{naze2020}
\begin{multline}
\Psi_0(t)=\exp{\left(-\gamma|t|\right)}\left[\frac{2\omega_0^2}{\omega^2}\right.\\
\left.+\left(\frac{\omega^2-2\omega_0^2}{\omega^2}\right)\cos{\omega t}+\frac{\gamma}{\omega}\sin{\omega |t|}\right],
\end{multline}
where $\omega=\sqrt{4\omega_0^2-\gamma^2}$. The coefficients $a_n$ will be
\be
a_{-2}=1,\quad a_{-1}=\tau_R=\frac{\gamma^2+\omega_0^2}{2\gamma\omega_0},\quad 
\ee
\be
a_{n}= \frac{1}{n!}\frac{\partial^{n}}{\partial s^{n}}\frac{(s+\gamma)(s^2+2\gamma s+4\omega_0^2)}{2s^2\omega_0^2(2\gamma+s)}\Big|_{s=0},
\ee
for $n\ge 0$, where $\tau_R$ is the relaxation timescale of the system. The optimal protocol will be
\be
g^*(t) = \frac{t+\tau_R}{\tau+2\tau_R}+\sum_{n=0}^{\infty}\frac{a_n (\delta^{(n)}(t)-\delta^{(n)}(\tau-t))}{\tau+2\tau_R}.
\ee

\subsection{Sinc relaxation function}

In Refs.~\cite{naze2023adiabatic,naze2023quantum}, when the time-averaged work was performed in a thermally isolated system, it behaves as performing an isothermal process with its typical characteristic time. In particular, for thermally isolated systems that have a relaxation function equal to
\be
\Psi_0(t)=\cos{(\omega t)},
\ee
such as the harmonic oscillator~\cite{acconcia2015degenerate}, Landau-Zener model~\cite{naze2023adiabatic} and statistical anyons~\cite{myers2021thermodynamics}, their time-averaged relaxation function is 
\be
\overline{\Psi}_0(t)=\text{sinc}\left(\omega t\right)=\text{sinc}\left(\frac{\pi}{2} \frac{t}{\tau_R}\right),
\ee
where $\tau_R$ is the relaxation timescale of the new relaxation function of the system. By applying the method, the terms $a_n$ will be
\be
a_{-2}=1,\quad a_{-1}=\tau_R,
\ee
\be
a_{n}=\frac{1}{(n+2)!}\frac{\partial^{(n+2)}}{\partial s^{(n+2)}}\frac{1}{(s^2-2 \tau_R s^3 \,\text{acot}(2 \tau_R s/\pi)/\pi)}\Big|_{s=0},
\ee
for $n\ge 0$. The optimal protocol will be
\be
g^*(t) = \frac{t+\tau_R}{\tau+2\tau_R}+\sum_{n=0}^{\infty}\frac{a_n (\delta^{(n)}(t)-\delta^{(n)}(\tau-t))}{\tau+2\tau_R}.
\ee

\subsection{Gaussian relaxation function}

Another relaxation function that satisfies the criteria of compatibility with the Second Law of Thermodynamics~\cite{naze2020} is the Gaussian relaxation function
\be
\Psi_0(t)=\exp\left(-\frac{\pi}{4}\left(\frac{t}{\tau_R}\right)^2\right),
\ee
where $\tau_R$ is the relaxation timescale of the system. At this time, as far as I know, there is no physical system whose relaxation function is modeled by this Gaussian one. However, it is important to know if the method works if one finds a possible system where works the modeling by this relaxation function. By applying the method, the terms $a_n$ will be
\be
a_{-2}=1,\quad a_{-1}=\tau_R,
\ee
and
\be
a_{n}=\frac{1}{n!}\frac{\partial^{n}}{\partial s^{n}}\frac{1}{(s^2-\tau_R s^3 \exp(\tau_R^2 s^2/\pi)\text{Erfc}(\tau_R s/\sqrt{\pi})}\Big|_{s=0},
\ee
for $n\ge 0$. Here $\text{Erfc}(x)$ is the complementary error function. The optimal protocol will be
\be
g^*(t) = \frac{t+\tau_R}{\tau+2\tau_R}+\sum_{n=0}^{\infty}\frac{a_n (\delta^{(n)}(t)-\delta^{(n)}(\tau-t))}{\tau+2\tau_R}.
\ee

\subsection{Bessel relaxation function}

The Bessel relaxation function is given by
\be
\Psi_0(t)=J_0\left(\frac{t}{\tau_R}\right),
\ee
where $J_0$ is the Bessel function of the first kind with $\nu=0$ and $\tau_R$ is its relaxation timescale. It satisfies the criteria for compatibility with the Second Law of Thermodynamics~\cite{naze2020}. Such relaxation function can model the Ising chain subjected to a time-dependent magnetic field and evolving in time in a non-perturbed solution according to Glauber-Ising dynamics~\cite{glauber1963time}. By applying the method, the terms $a_n$ will be
\be
a_{-2}=1,\quad a_{-1}=\tau_R,
\ee
and
\be
a_{n}=\frac{1}{n!}\frac{\partial^{n}}{\partial s^{n}}(s^{-2}+\tau_R^2+\tau_R s^{-1}\sqrt{1+\tau_R^2 s^2})\Big|_{s=0},
\ee
with $n\ge 0$. The optimal protocol will be
\be
g^*(t) = \frac{t+\tau_R}{\tau+2\tau_R}+\sum_{n=0}^{\infty}\frac{a_n (\delta^{(n)}(t)-\delta^{(n)}(\tau-t))}{\tau+2\tau_R}.
\ee

\section{Discussion}

\subsection{Continuous part}

For all examples treated here, the continuous part of the optimal protocol $g^*_C(t)$ was given by
\be
g^*_{C}(t)=\frac{t+\tau_R}{\tau+2\tau_R}.
\ee
Such a solution is consistent with previous results for slowly-varying and sudden cases where respectively for the asymptotic limits $\tau\gg\tau_R$ and $\tau\ll\tau_R$~\cite{naze2022optimal}
\be
\lim_{\tau\gg\tau_R}g^*_C(t)=\frac{t}{\tau},\quad \lim_{\tau\ll\tau_R}g^*_C(t)=\frac{1}{2}.
\ee
Also, this continuous part presents an upper and lower bounds
\be
0\le g^*_C(t)\le 1,
\ee
for all $t\in[0,\tau]$, $0\le\tau/\tau_R\le\infty$, meaning that it fulfills the linear-response theory hypothesis $|g(t)\delta\lambda/\lambda_0|\ll 1$.

\subsection{Singular part}

For all examples treated here, the singular part of the optimal protocol $g^*_S(t)$ was given by
\be
g^*_S(t)=\sum_{n=0}^{\infty}\frac{a_n (\delta^{(n)}(t)-\delta^{(n)}(\tau-t))}{\tau+2\tau_R}.
\ee
For the asymptotic limits $\tau\gg\tau_R$ and $\tau\ll\tau_R$, one has
\be
\lim_{\tau\gg\tau_R}g^*_S(t)=\sum_{n=0}^{\infty}\frac{a_n (\delta^{(n)}(t)-\delta^{(n)}(\tau-t))}{\tau},
\ee
\be
\lim_{\tau\ll\tau_R}g^*_S(t)=\sum_{n=0}^{\infty}\frac{a_n (\delta^{(n)}(t)-\delta^{(n)}(\tau-t))}{2\tau_R}.
\ee
Since Dirac deltas and their derivatives can be interpreted as singular points, will they be outside the hypothesis of linear response theory, where $|g(t)\delta\lambda/\lambda_0|\ll 1$? This question is not so simple to answer, mainly because such a criterion is just a heuristic one. Indeed, to answer this in a complete manner, it is necessary to know their non-linear response in second-order expansion and make a direct comparison with linear response theory. However, at this point, this is an open question, even for continuous protocols. 

Another important question is: what are these Dirac deltas and derivatives in practice? They are nothing more than impulse-like terms. In practice, the Dirac deltas can be approximated by using very sharp Gaussian functions. Their derivatives can be implemented in principle by using the property $t^n\delta^{(n)}(t)=(-1)^n\delta(t)$, transforming them into a Dirac delta. Also, there are timescales involved in each impulse-like term, showing that they have a physical reality. Indeed, considering
\be
\tau_I^n=\left(\frac{a_n}{\tau_R}\right)^{\frac{1}{n+1}},
\ee
one has
\be
g^*_S(t)=\sum_{n=0}^{\infty}\frac{\tau_R (\delta^{(n)}(t/\tau_I^n)-\delta^{(n)}((\tau-t)/\tau_I^n))}{\tau+2\tau_R},
\ee
where the derivatives are taken on $t/\tau_I^n$. Therefore, to achieve optimality three protocol types must be considered: continuous-, jump-, and impulse-like terms.

\subsection{Whole protocol}

Considering the whole protocol, in the asymptotic limit $\tau\gg\tau_R$ it is
\be
\lim_{\tau\gg\tau_R}g^*(t)=\frac{t}{\tau}+\sum_{n=0}^{\infty}\frac{a_n (\delta^{(n)}(t)-\delta^{(n)}(\tau-t))}{\tau},
\ee
while in the asymptotic limit $\tau\ll\tau_R$ in
\be
\lim_{\tau\ll\tau_R}g^*(t)=\frac{1}{2}+\sum_{n=0}^{\infty}\frac{a_n (\delta^{(n)}(t)-\delta^{(n)}(\tau-t))}{2\tau_R}.
\ee
Observe that our demonstration in Ref.~\cite{naze2022optimal} for the optimal protocol in those asymptotic limits did not consider the existence of Dirac deltas. However, as these limiting cases are solution of Euler-Lagrange equation, the demonstration presented there refers only to the case where $a_n=0$, for $n\ge 0$.

\subsection{Optimal irreversible work and variance}

Finally, the optimal irreversible work is
\be
\begin{split}
W_{\rm irr}^*(\tau) &=\ \frac{\delta\lambda^2}{2}\Psi_0(0)+\frac{\delta\lambda^2}{2}\int_0^\tau \dot{\Psi}_0(\tau-t)g^*_C(t)dt\\
&+\frac{\delta\lambda^2}{2}\sum_{n=0}^{\infty}\frac{a_{2n}[\dot{\Psi}_0^{(2n)}(\tau)-\dot{\Psi}_0^{(2n)}(0)]}{\tau+2\tau_R}\\
&-\frac{\delta\lambda^2}{2}\sum_{n=0}^{\infty}\frac{a_{2n+1}[\dot{\Psi}_0^{(2n+1)}(\tau)-\dot{\Psi}_0^{(2n+1)}(0)]}{\tau+2\tau_R},
\label{eq:optwirr}
\end{split}
\ee
whose approximation in the asymptotic limit $\tau\gg\tau_R$ is
\be
\begin{split}
\lim_{\tau\gg\tau_R}W_{\rm irr}^*(\tau)&=\frac{\delta\lambda^2}{2\tau}\int_0^\tau \Psi_0(t)dt\\
&+\frac{\delta\lambda^2}{2}\sum_{n=0}^{\infty}\frac{a_{2n}[\dot{\Psi}_0^{(2n)}(\tau)-\dot{\Psi}_0^{(2n)}(0)]}{\tau}\\
&-\frac{\delta\lambda^2}{2}\sum_{n=0}^{\infty}\frac{a_{2n+1}[\dot{\Psi}_0^{(2n+1)}(\tau)-\dot{\Psi}_0^{(2n+1)}(0)]}{\tau},
\label{eq:optwirr}
\end{split}
\ee
where the scaling with $\tau^{-1}$ agrees with previous study~\cite{sivak2012}. Also, the optimal irreversible work vanishes as expected by the Second Law of Thermodynamics in the limit $\tau\rightarrow\infty$. According to Ref.~\cite{naze2022optimal}, such a result is achievable only with the continuous part of the solution, showing that the singular part is not relevant in such a regime. In the asymptotic limit $\tau\ll\tau_R$
\be
\begin{split}
\lim_{\tau\ll\tau_R}W_{\rm irr}^*(\tau)&=\frac{\delta\lambda^2}{4}\Psi_0(0)+\frac{\delta\lambda^2}{4}\Psi_0(\tau)\\
&+\frac{\delta\lambda^2}{2}\sum_{n=0}^{\infty}\frac{a_{2n}[\dot{\Psi}_0^{(2n)}(\tau)-\dot{\Psi}_0^{(2n)}(0)]}{2\tau_R}\\
&-\frac{\delta\lambda^2}{2}\sum_{n=0}^{\infty}\frac{a_{2n+1}[\dot{\Psi}_0^{(2n+1)}(\tau)-\dot{\Psi}_0^{(2n+1)}(0)]}{2\tau_R},
\label{eq:optwirr}
\end{split}
\ee
which curiously shows a dependence on $\tau$. For $\tau\rightarrow 0^+$, one has
\be
\lim_{\tau\rightarrow 0^+}W_{\rm irr}^*=\frac{\delta\lambda^2}{2}\Psi_0(0),
\label{eq:optwirr}
\ee
showing that the irreversible work is a non-null constant, as expected by a rapidly-driven protocol where the sudden change in the parameter irreversibly dissipates energy~\cite{blaber2021steps,rolandi2023optimal}. Observe also that such optimal irreversible work is the same as the one calculated in Ref.~\cite{naze2022optimal} considering only jumps at the beginning and end of the process. This illustrates that the singular part has no effect in optimizing the irreversible work in this particular regime.

The optimal work variance can be obtained by using the same optimal protocol, which is given according to Eq.~\eqref{eq:fdr} by
\be
\begin{split}
\sigma_{W}^{2*}(\tau) &=\ \frac{\delta\lambda^2}{\beta}\Psi_0(0)+\frac{\delta\lambda^2}{\beta}\int_0^\tau \dot{\Psi}_0(\tau-t)g^*_C(t)dt\\
&+\frac{\delta\lambda^2}{\beta}\sum_{n=0}^{\infty}\frac{a_{2n}[\dot{\Psi}_0^{(2n)}(\tau)-\dot{\Psi}_0^{(2n)}(0)]}{\tau+2\tau_R}\\
&-\frac{\delta\lambda^2}{\beta}\sum_{n=0}^{\infty}\frac{a_{2n+1}[\dot{\Psi}_0^{(2n+1)}(\tau)-\dot{\Psi}_0^{(2n+1)}(0)]}{\tau+2\tau_R},
\label{eq:optvar}
\end{split}
\ee
which the asymptotic limits $\tau\gg\tau_R$ and $\tau\ll\tau_R$ are given similarly. The dependence on the initial temperature shows that the variance increases as great as it is. Although reasonable, such a feature has not been explored in the few but important experiments carried out in optimal control~\cite{loos2024universal}.

\subsection{Discussion about numerical comparison}

Although the results deduced in this work illustrate the analytical method, a numerical comparison of those examples will corroborate the optimality of the solution. Reference~\cite{naze2023global} informs us that any method that finds local minima is enough to find global minima. However, such a method must be able to produce solutions in the space of distributions. Genetic programming~\cite{koza1992,duriez2016} seems to fail in such a sense since its solutions depend on a function basis that is composed of continuous functions. Therefore, it is expected that the numerical solutions obtained by this method will be non-optimal. However, the method used in Ref.~\cite{deffner2018} produces at least Dirac deltas, although producing their derivatives seems to be a large step. Future research will be done in this direction to explore those raised points.

\section{Final remarks}
\label{sec:final} 

In this work, I presented a method to solve the Euler-Lagrange integral equation that furnishes the optimal protocol to minimize the irreversible work and its variance in the context of finite-time, isothermal and weakly driven processes. It relies only on the time-reversal symmetric property of the optimal protocol, which has been proved. The solution is composed of a straight line with jumps at the beginning and end of the process, and impulse-like terms of Dirac deltas and their derivatives. The appearance of such distribution terms is central since numerical investigations might not be able to achieve such optimal solutions if the admissible functions remain in real space. Already known results are recovered and many new examples are solved corroborating the result. The slowly-varying and sudden cases are deduced as well. A numerical comparison to corroborate the optimality of the universal optimal protocol requires numerical solutions in the space of distributions, which are not available with the methods present in the literature. Finally, verifying such generalization opens the door to whether this aspect remains for higher orders than linear response theory.   

\bibliography{ASLR}

\begin{thebibliography}{34}
\bibitem [{hrm()}]{hrmethod}%
  \BibitemOpen
  \href@noop {} {}\bibinfo {note} {Horner-Ruffini polynomial division method
  finds a function (quotient) such that, when multiplied by a polynomial
  (divisor), furnishes another function (dividend). In the case treated,
  $1/s^2$ is the dividend.}\BibitemShut {Stop}%
\end{thebibliography}

\end{document}